\documentclass[pre,aps,twocolumn,superscriptaddress]{revtex4-1}
\pdfoutput=1
\usepackage{amssymb}
\usepackage{amsmath}
\usepackage{subfigure}
\usepackage{graphicx}
\usepackage{textcomp}
\usepackage{url}
\usepackage{color}
\usepackage{cases}
\usepackage[normalem]{ulem}
\usepackage{float}

\usepackage[colorlinks=true, urlcolor=blue, anchorcolor=blue, citecolor=blue,filecolor=blue,linkcolor=blue,menucolor=blue
]{hyperref}

\usepackage[titletoc,title]{appendix}

\usepackage{color}

\graphicspath{{data/}{figures/}}

\begin{document}
\title{First-passage area distribution and optimal fluctuations of
fractional Brownian motion}

\author{Alexander K. Hartmann}
\email{a.hartmann@uni-oldenburg.de}
\affiliation{Institut f\"{u}r Physik, Universit{\aa}t Oldenburg - 26111 Oldenburg, Germany}
\author{Baruch Meerson}
\email{meerson@mail.huji.ac.il}
\affiliation{Racah Institute of Physics, Hebrew University of
Jerusalem, Jerusalem 91904, Israel}


\begin{abstract}
We study the probability distribution $P(A)$ of the area $A=\int_0^T x(t) dt$ swept under fractional Brownian motion (fBm) $x(t)$ until its first passage time $T$ to the origin. The process starts at $t=0$ from a specified point $x=L$. We show that $P(A)$ obeys exact scaling relation
$$
P(A) = \frac{D^\frac{1}{2H}}{L^{1+\frac{1}{H}}}\,\Phi_H\left(\frac{D^\frac{1}{2H} A}{L^{1+\frac{1}{H}}}\right)\,,
$$
where $0<H<1$ is the Hurst exponent characterizing the fBm,
$D$ is the coefficient of fractional diffusion, and $\Phi_H(z)$ is a scaling function. The small-$A$ tail of $P(A)$ has been recently predicted by Meerson and Oshanin [Phys. Rev. E \textbf{105}, 064137 (2022)], who showed that it has an essential singularity at $A=0$, the character of which depends on $H$.  Here we
determine the large-$A$ tail of $P(A)$. It is a fat tail, in particular such
that the average value of the first-passage area $A$ diverges for all $H$.
We also verify the predictions
for both tails by performing simple-sampling as well as
large-deviation Monte Carlo simulations. The verification includes measurements of $P(A)$ up to probability densities as small as $10^{-190}$.
We also perform direct observations of paths conditioned on the area $A$.
For the steep small-$A$ tail of $P(A)$  the ``optimal paths'', \textit{i.e.} the
most probable trajectories of the fBm, dominate the statistics.
Finally, we discuss extensions of theory to a more general first-passage
functional of the fBm.
\end{abstract}

\maketitle


\section{Introduction}
\label{intro}
The fractional Brownian motion (fBm), introduced by Kolmogorov \cite{Kolmogorov} and by Mandelbrot and van Ness \cite{Mandelbrot}, is a
non-Markovian Gaussian stochastic process which has found
applications across many disciplines, from anomalous diffusion in cellular environments
to mathematical finance. These include dynamics of stochastic interfaces \cite{Krug}, particle
transport in crowded fluids \cite{weiss,weiss2}, sub-diffusion of bacterial loci in a cytoplasm \cite{weber},
telomere diffusion in
the cell nucleus \cite{garini,prx},
modeling of conformations of serotonergic axons \cite{vojta},
dynamics of tagged beads of polymers \cite{Walter,Amitai},
translocation of a polymer through a pore \cite{Amitai,Zoia,Dubbeldam,Palyulin}, single-file
diffusion in ion channels \cite{Kukla,Wei,chanel}, \textit{etc.} A review can be found in Ref.~\cite{ralf}.

Let $x(t)$ denote a realization, such that $x(0)=0$, of a one-dimensional fBm. Being a Gaussian process with zero mean, the fBm is completely defined by its two-time covariance. For the one-sided process, defined on the interval $0\leq t<\infty$,
the covariance is given by \cite{Kolmogorov,Mandelbrot}
\begin{equation}\label{kappa1}
\kappa(t,t')\!\equiv\!\langle x(t) x(t')\rangle\!=\!
D\left(t^{2H}+t'^{2H}-|t-t'|^{2H}\right)\,,
\end{equation}
where $H \in (0,1)$  is the Hurst exponent, and $D=\text{const}>0$ is the coefficient of fractional diffusion. The fBm describes a family of anomalous diffusion.  The mean-square displacement
$\langle x^2(t)\rangle = 2 D t^{2H}$ 
grows sublinearly with time for $H < 1/2$ (subdiffusion) and superlinearly for $H > 1/2$ (superdiffusion). In the borderline case $H = 1/2$ the standard diffusion is recovered.

\begin{figure} [ht]
\includegraphics[width=0.3\textwidth,clip=]{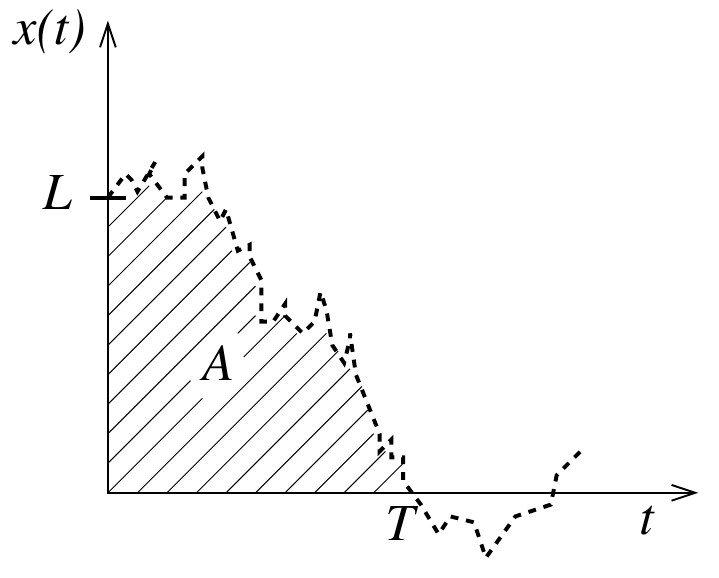}
\caption{Example of a random process, starting at $x(0)=L$ and hitting
$x=0$ for the first time at time $T$. The area under the curve
until time $T$ is denoted by $A$, and its distribution for the fBm
is the focus of this work.}
\label{fig:sample_path}
\end{figure}

Due to its non-Markovian nature, the fBm presents a significant challenge for theorists. Statistics of various random quantities, which  for the standard Brownian motion were determined long ago, remain unknown for the fBm. Here we will study the fBm until its first passage to a
given point $x=L$. Equivalently, we can transform the coordinate, $x \to L-x$, so that the starting point becomes $x=L$, and the first-passage point is $x=0$.
In particular, we study
the probability distribution $P(A)$ of the first-passage area of fBm,
\begin{equation}\label{fparea}
A=\int_0^T x(t) \,dt\,,
\end{equation}
where $T$ is the first-passage time with $x(T)=0$,   see
Fig.~\ref{fig:sample_path} for an illustrative example.
The first-passage area $A$ is a random quantity because of
two reasons: (i) the random character of the paths $x(t)$, and (ii)
the randomness of $T$ itself.

For the standard Brownian motion, $H=1/2$, the probability distribution $P(A)$ is known  \cite{KM2005,MM2020a}:
\begin{equation}\label{PAexact}
P(A) = \frac{L}{3^{2/3} \Gamma
   \left(1/3\right) (D A^4)^{1/3}}\, \exp\left(-\frac{L^3}{9 D A}\right)\,,
   \end{equation}
where $\Gamma(\dots)$ is the gamma function. Noticeable is the steep decrease and the essential singularity of $P(A)$ at $A\to 0$, and a fat tail $P(A)\sim A^{-4/3}$ at $A\to \infty$. Because of this tail, the average value of $A$ is infinite.

The first-passage area of the standard Brownian motion appears in many applications:
from combinatorics and queueing theory to the statistics of avalanches in self-organized
criticality~\cite{KM2005}. In the context of the queueing theory, $x(t)$ may represent
the length of a queue in front of a ticket counter during a busy period, whereas
$\mathcal A$ is the total serving time of the customers during the busy period.
The first-passage area also appears in the study of the distribution of avalanche
sizes in the directed Abelian sandpile model~\cite{DR1989,K2004},
of the area of staircase polygons in compact directed percolation~\cite{PB1995,C2002,K2004} and
of the collapse time of a ball bouncing on a noisy platform~\cite{MK2007}. Extending the underlying model of stochastic processes from the standard Brownian motion to fBm would make it possible to account for non-Markovianity but still keep the important properties of dynamical scale invariance, stationarity of the increment, and Gaussianity.

The probability distribution $P(A)$ of the fBm depends on the dimensional quantities  $A$, $L$ and $D$ and on the dimensionless parameter $H$. A straightforward dimensional analysis (see, \textit{e.g.} Ref. \cite{Barenblatt}) yields exact
scaling behavior of $P(A)$:
\begin{equation}\label{exactscaling}
P(A) = \frac{D^\frac{1}{2H}}{L^{1+\frac{1}{H}}}\,\Phi_H\left(\frac{D^\frac{1}{2H} A}{L^{1+\frac{1}{H}}}\right)\,.
\end{equation}
The scaling function $\Phi_H(z)$ is presently unknown except for $H=1/2$, see Eq.~(\ref{PAexact}), where $\Phi_{1/2}(z)\sim z^{-\frac{4}{3}} e^{-\frac{1}{9z}}$.
The maximum value of $P(A)$, of order $D^{\frac{1}{2H}}/L^{1+\frac{1}{H}}$, is reached at
\begin{equation}\label{Ascaling}
A\sim L^{1+\frac{1}{H}} D^{-\frac{1}{2H}}\,.
\end{equation}
Equation~(\ref{Ascaling}) describes the typical scaling of the first-passage area with $L$ and $D$, and it has a simple explanation. Indeed, $x(t)$ exhibits the anomalous diffusion scaling $x(t) \sim \sqrt{D} \,t^H$. As a result, $A(T)=\int_0^T x(t) dt$ scales as $\sqrt{D} \,T^{H+1}$. In its turn, $T$ scales as $(L/\sqrt{D})^{1/H}$, leading to the scaling behavior (\ref{Ascaling}).

The small-$z$ asymptotic of  $\Phi_H(z)$ was recently evaluated in Ref. \cite{MO2022} for any $0<H<1$. Up to an unknown pre-exponential factor, it is the following:
\begin{equation}\label{smallz}
\Phi_H(z\to 0) \sim \exp\left[-\sigma(H) z^{-2H}\right]\,,
\end{equation}
where \cite{MO2022}
\begin{numcases}
{{\sigma(H)} =}\!\!\frac{1}{2^{2+2H}(1-H^2)(1+H)^{2 H}},& \!\!$H \leq 1/2\,.$ \label{sigmaless} \\
 \!\frac{H^{2H}}{2 (1+H)^{1+2 H}},& \!\!$H\geq 1/2\,.$
\label{sigmamore}
\end{numcases}
As a result,
\begin{equation}\label{smallA}
P(A\to 0) \sim \exp\left[-\sigma(H) \frac{L^{2+2H}}{DA^{2H}}\right]\,.
\end{equation}
This steep small-$A$ tail exhibits an essential singularity at $A=0$, the character of which is determined by the Hurst exponent $H$.

The $A\to 0$ tail was determined in Ref. \cite{MO2022} by using the optimal-fluctuation method -- essentially geometrical optics of the fBm. The method is based on the  determination of the optimal path -- the most likely realization of $x(t)$ which dominates $P(A)$ at specified value of $A$ and enables one to perform a saddle-point evaluation of the proper path integral of the fBm \cite{MBO2022}. The calculation boils down to a minimization of the Gaussian action of the fBm, obeying the boundary conditions $x(L)=L$ and $x(T) = 0$ and constrained by Eq.~(\ref{fparea}) and by the inequality $x(t)>0$ for all $0<t<T$. The minimization is performed with respect both to the path $x(t)$, and to the first-passage time $T$ \cite{MO2022}. The minimization leads to a linear integral equation -- a non-local generalization of the Euler-Lagrange equation for local functional -- for the optimal path $x(t)$.

In addition to presenting the exact scaling behavior of $P(A)$ in Eq.~(\ref{exactscaling}), here we
extend the previous study \cite{MO2022} in two directions. First, we offer a simple scaling argument which predicts, up to a factor $O(1)$ which can depend only on $H$,
the $A\to \infty$ tail of the probability distribution $P(A)$:
\begin{equation}\label{largeA}
P(A\to \infty) \sim \frac{L^{\frac{1}{H}-1}}{D^{\frac{1-H}{2H(H+1)}}A^{\frac{2}{H+1}}}\,.
\end{equation}
This fat tail, with an $H$-dependent exponent,
corresponds to the large-$z$ asymptotic
\begin{equation}\label{largez}
\Phi_H(z\to \infty) \sim z^{-\frac{2}{H+1}}\,,
\end{equation}
of the scaling function  $\Phi_H(z)$ entering Eq.~(\ref{exactscaling}).   For $H=1/2$ Eq.~(\ref{largeA}) predicts an $A^{-4/3}$ tail in agreement with the exact result ~(\ref{PAexact}).

Second, we perform simple-sampling as well as large-deviation Monte-Carlo simulations of fractional random walks in order to test the theoretical predictions  of the scaling behavior (\ref{exactscaling}) and of the tails (\ref{smallA}) and (\ref{largeA}).
We also directly observe in the simulations the paths $x(t)$ corresponding to the specified $A$.
For large $A$, as described by the fat distribution tail (\ref{largeA}),
we find that, as to be expected,
multiple paths contribute to the statistics. On the contrary,
for small $A$, corresponding to the tail~(\ref{smallA}), the behavior is dominated
by the optimal paths which we
compare with those predicted by geometrical optics of the fBm \cite{MO2022}.

\section{Large-$A$ tail of the first-passage area distribution}

The large-$A$ tail is contributed to by multiple paths. Their common feature is that the particle spends a very long time, in comparison with the characteristic fractional diffusion time $(L/\sqrt{D})^{1/H}$, before reaching the origin for the first time. For such paths, biased by a very large $A$, a typical deviation of the particle from the origin, $\ell \sim \sqrt{D} \,t^{H}$, is much larger than $L$. As a result, the typical area $A$, swept by $x(t)$ until the first-passage time $T$, scales as
\begin{equation}\label{Atyp}
A \sim \sqrt{D}\, T^{H+1}
\end{equation}
and, to a leading order, is independent of $L$.  Now we use the probability distribution of the first-passage time $f(T)$ and make a change of variables from $T$ to $A$ using Eq.~(\ref{Atyp}).
The complete first-passage time distribution $f(T)$ of the fBm has not been determined as yet. However, its small-$T$ and large-$T$ tails are available. The small-$T$ tail, $f(T\to 0) \sim \exp\left(-L^2/4D T^{2H}\right)$,
has been found by a geometrical-optics calculation \cite{MO2022}, but for the purpose of evaluating the large-$A$ tail of $P(A)$ we need the large-$T$ tail of $f(T)$. This tail,
\begin{equation}\label{flargeA}
f(T\to \infty) \sim \frac{L^{\frac{1}{H}-1}}{D^{\frac{1-H}{2H}}T^{2-H}}\,,
\end{equation}
was conjectured in Refs. \cite{Hansen,Maslov,Ding,Krugetal}, verified in numerical simulations \cite{Ding,Krugetal} and ultimately proved rigorously \cite{Molchan,Aurzada}. Using the relation
\begin{equation}\label{ftoP}
P(A)\,dA = f(T)\,dT
\end{equation}
and Eqs.~(\ref{Atyp}) and (\ref{ftoP}), we arrive at the scaling behavior of the large-$A$ tail of $P(A)$, announced in Eq.~(\ref{largeA}).

\section{Simulation methods}

We approximate the fBm by a $K$-step walk $x(t_l)$ ($l=0,1,\ldots,K$, where $K\gg 1$)
at discrete times $t_0,t_1,\ldots$
with step size $\Delta t= t_{l+1}-t_{i}=1$, \textit{i.e.} $t_l=l$. In order
to generate walks obeying the correlations (\ref{kappa1}) we note that
for the increments $\Delta x_l=x_{l+1} $ we obtain
\begin{equation}
C(m)\equiv \langle \Delta x_{l+m} \Delta x_l \rangle =
D(|m+1|^{2H}-2|m|^{2H}+ |m-1|^{2H})
\end{equation}
independent of $l$. To generate random increments which obey these correlations,
we apply an algorithm in the spirit of the Davies-Harte approach
\cite{davies1987}, actually
the {\em circulant embedding method}
 proposed in Refs. \cite{wood1994,dietrich1997} for a fast generation
of correlated random numbers. It is based on the application
of the fast Fourier transform (FFT)
to generate a longer periodic walk with $K'\ge 2K$ steps exhibiting
a symmetric correlation, \textit{i.e.} $C(l)$ for $l<K'/2$ and $C(K'-l)$ for $l\ge K'/2$.
Since FFT is used, it makes sense to choose $K'$ as a power of 2.
For the actual walks we then
take at most the first half of the generated increments,
so the periodicity does not play a  role. The method works by first
 generating $K'$ Gaussian random
numbers $\xi=(\xi_0,\xi_1,\ldots,\xi_{K'-1})$ with zero mean and unit variance.
These random numbers are then multiplied by a suitably scaled pre-computed
Fourier
transform of the desired correlation. The result is finally transformed back
into real space by an inverse FFT. For details see
Refs.~\cite{wood1994,fBm_MC2013}.
This yields the actual correlated increments $\Delta x_1, \ldots, \Delta x_k$
which are summed up to generate a walk starting at $x(0)=L$ by setting
\begin{equation}
x(l)=L+\sum_{j=1}^l \Delta x_l\,.
\end{equation}
Since the FFT runs in
$O(K' \ln K')$ time, and all other computations are performed in a linear time,
the approach is very efficient. For the FFT we used here the GNU scientific
library (GSL) \cite{gsl2006}.
This approach has been previously applied,
\textit{e.g.}, to generate fractional Brownian walks with absorbing boundary
conditions \cite{fBm_MC2013}.

We determine an approximate first-passage time to the target at $x=0$
by first finding the minimum
step $l_{\rm fp}$ where the walk becomes negative, \textit{i.e.}
\begin{equation}
l_{\rm fp} = \min \{l| x(l) <0 \}\,
\end{equation}
which might yield no result. In this case the walk does not exhibit
a first passage within $K$ steps and does not contribute to the
statistics, so it is discarded.

We denote the fraction of non-discarded walks, \textit{i.e.}
those which reached the target $x=0$, as $p_{\rm fp}$. For those walks,
we determine the actual first-passage time
$T$ by linearly interpolating between steps $l_{\rm fp}-1$ and $l_{\rm fp}$, \textit{i.e.}
\begin{equation}
T= l_{\rm fp}-1 + \frac{x(l_{\rm fp}-1)}{x(l_{\rm fp}-1)-x(l_{\rm fp})}\,.
\end{equation}
Note that this is an approximation of the first-passage
time due to the walk being a discrete approximation of fBm.
If one wanted to estimate the first-passage time $T$ of fBm with a higher
accuracy, one should use an iterative algorithm \cite{walter2020}, which
is based on refining the time step adaptively in regions which are
close to the target. Here, we are mostly interested in the first-passage area, so that a refinement just near the first-passage point
would not increase the accuracy much. Therefore, we use the trapezoidal rule
to estimate the area as
\begin{equation}
A = \sum_{l=1}^{l_{\rm fp}-1} \frac{x(l-1)+x(l)}{2}
+ (T-l_{\rm fp}+1)\frac{x(l_{\rm fp}-1)}{2}\,,
\end{equation}
where the final contribution is the triangle obtained from the last position
$x(l_{\rm fp}-1)$ before the target is reached until the estimated
first-passage time $T$ where $x(T)=0$ holds.

For a given value of $L$, one can generate many independent
vectors $\xi$ of Gaussian numbers,
obtain the resulting walks $x(l)$  and each time
calculate the corresponding area $A=A(\xi)$ for those walks which
pass $x=0$, as described above.
By measuring a histogram, properly
normalized such that the integral results in the fraction $p_{\rm fp}$ of paths
that exhibit a first passage, an estimate of $P(A)$ is obtained. By generating
$n$ walks within this \emph{simple-sampling} approach,
the distribution can be estimated down to probabilities $O(1/n)$,
\textit{e.g.}, $P\sim 10^{-6}$ for $10^6$ walks.  For the small-$A$
tail of the distribution, however, we want to verify the theoretical predictions in regions
where the probability densities are much smaller.

For this reason, we also
employ a large-deviation approach which allows one to access the tails
of the distributions. Such approaches have a long history in the
field of variance-reduction techniques \cite{hammersley1956}, and they were
introduced to physics, \textit{e.g.} in transition-path sampling \cite{dellago1998,crooks2001}.
This and other approaches have been applied in physics
to many different problems,
\textit{e.g.}, random graph properties \cite{hartmann2004, hartmann2011, hartmann2017, schawe2019}, the resilience of power grids \cite{Dewenter_2015, feld19},
random walks \cite{claussen2015, Schawe2019No2, schawe2020, Chevallier2020},
ground states of Ising spin glasses \cite{Koerner2006},
longest increasing subsequences  \cite{boerjes2019,krabbe2020},
and the Kardar-Parisi-Zhang equation \cite{Hartmann2018,HMS1,HMS2}.
For a review of the general techniques see Ref.~\cite{bucklew2004}.

To obtain realizations of the paths which correspond to extremely small values of
the area $A$,
we do not sample the random number $\xi$  according to its natural
Gaussian product weight $G(\xi)$,
but according to the modified weight
$Q_{\Theta}(\xi)\sim G(\xi)\exp(-A(\xi)/\Theta)$, \textit{i.e.} with an
exponential \emph{bias}. $\Theta$ is an
auxiliary ``temperature'' parameter, which allows us to shift the resulting
 distributions of the area $A$, \textit{e.g.},
to smaller than typical values
by using $\Theta>0$ close to zero.
Note that $\Theta=\infty$ corresponds to the original statistics.
 With known algorithms, the vector $\xi$ of random numbers  cannot
be generated directly according to the modified weight $Q_{\Theta}(\xi)$.
Instead we use a standard Markov-chain approach with the
Metropolis-Hastings algorithm \cite{newman1999},
where the configurations of the Markov chain are
the vectors $\xi$ of random numbers. Since each vector $\xi$ corresponds
to a walk and therefore to an area $A$, one obtains a chain of sampled
values of $A$, depending on $\Theta$ and on the
the physical parameters  $L$, $D$
and $H$, which we do not include in the notation.
By sampling for different values
of $\Theta$ one obtains different distributions $P_{\Theta}(A)$ which
are centered about different typical values $A_{\rm typ}(\Theta)$.
By combining
and normalizing the distributions jointly for the different values of $\Theta$,
one obtains $P(A)$ over a large range of the support
\cite{Hartmann2018}, down to probabilities as small as $10^{-100}$.
For details of the approach
see Refs. \cite{align2002,work_ising2014,Hartmann2018}.
Here we extend it by storing
during the simulation (after equilibration) configurations $\xi$
 and the corresponding walks $x(t)$ at various
values of $\Theta$. This allows us to further analyze the walks, also
conditioned on the value of the area $A$.

\section{Theory versus simulations}

To compare the analytical predictions with numerical results and analyze the
walk paths, we performed simulations for $H=1/4$, $1/2$ and $3/4$, which represent
anticorrelated, uncorrelated,
and positively correlated increments, respectively. Without loss of generality we set $D=1$.
We worked with selected values of
the initial distance $L\le 100$, which were chosen to be sufficiently large so that the fractional random walk is a good approximation to the fBm, but not too large so that the first-passage probability
$p_{\rm fp}$ is not too small.

In our simple-sampling simulations we generated a large number of walks and measured
their properties. The longest walks that we considered had $K=10^{7}$ steps. For each combination of the parameters, we sampled at least $10^6$ walks. Sometimes, for smaller $K$, we sampled
more than $10^7$ walks.

To access the steep small-$A$ tail of the distribution $P(A)$, predicted by Eq.~(\ref{smallA}),
we used the large-deviation approach. Here much shorter walks,
consisting of at most $K=10^5$ steps, turned out to be sufficient.
We considered several values of the ``temperature'' parameter
$\Theta\in[1/2,10^6]$.
The largest number $n_{\Theta}$ of different values of $\Theta$, needed
to obtain the desired distribution $P(A)$, was
$n_{\Theta}=17$ for $H=1/4$ and $L=50$.

Since the large-$A$ tail of $P(A)$ is expected to follow a
slowly-decreasing power law, see Eq.~(\ref{largeA}), large-deviation simulations here would require prohibitively
large walk lengths. Here, the simple sampling results of sufficiently long walks turned out to be sufficient for the observation of the predicted behavior of the tail.

\subsection{First-passage area distribution $P(A)$}

\begin{figure} [ht]
\includegraphics[width=0.40\textwidth,clip=]{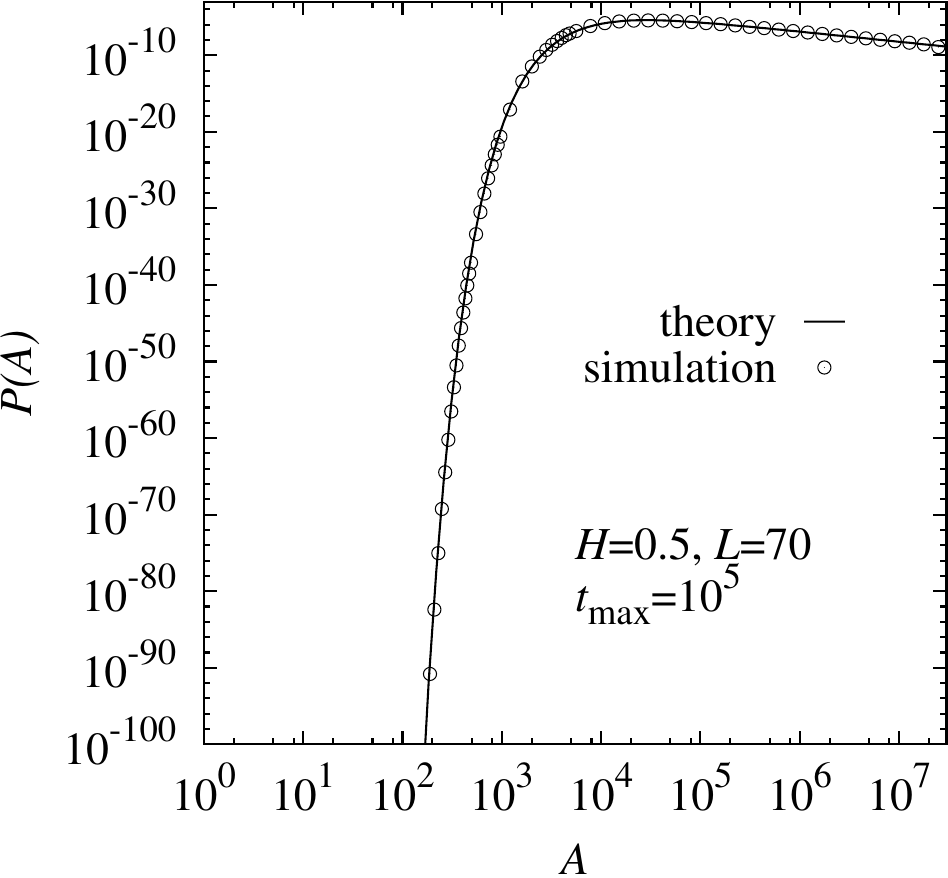}
\caption{Simulated first-passage area distribution $P(A)$ for the standard random walk versus exact theoretical prediction (\ref{PAexact}) for the standard Brownian motion ($H=1/2$) with $L=70$.}
\label{test05}
\end{figure}

As a validation, we measured in the simulations the first-passage area distribution $P(A)$ for the standard random walk $H=1/2$ with $L=70$ and compared it with the exact result (\ref{PAexact}) \cite{KM2005,MM2020a} for the standard Brownian motion, $H=1/2$. This comparison is shown in Fig. \ref{test05}, and a very good agreement is observed, showing the the large-deviation approach works very well.

\begin{figure} [ht]
\includegraphics[width=0.40\textwidth,clip=]{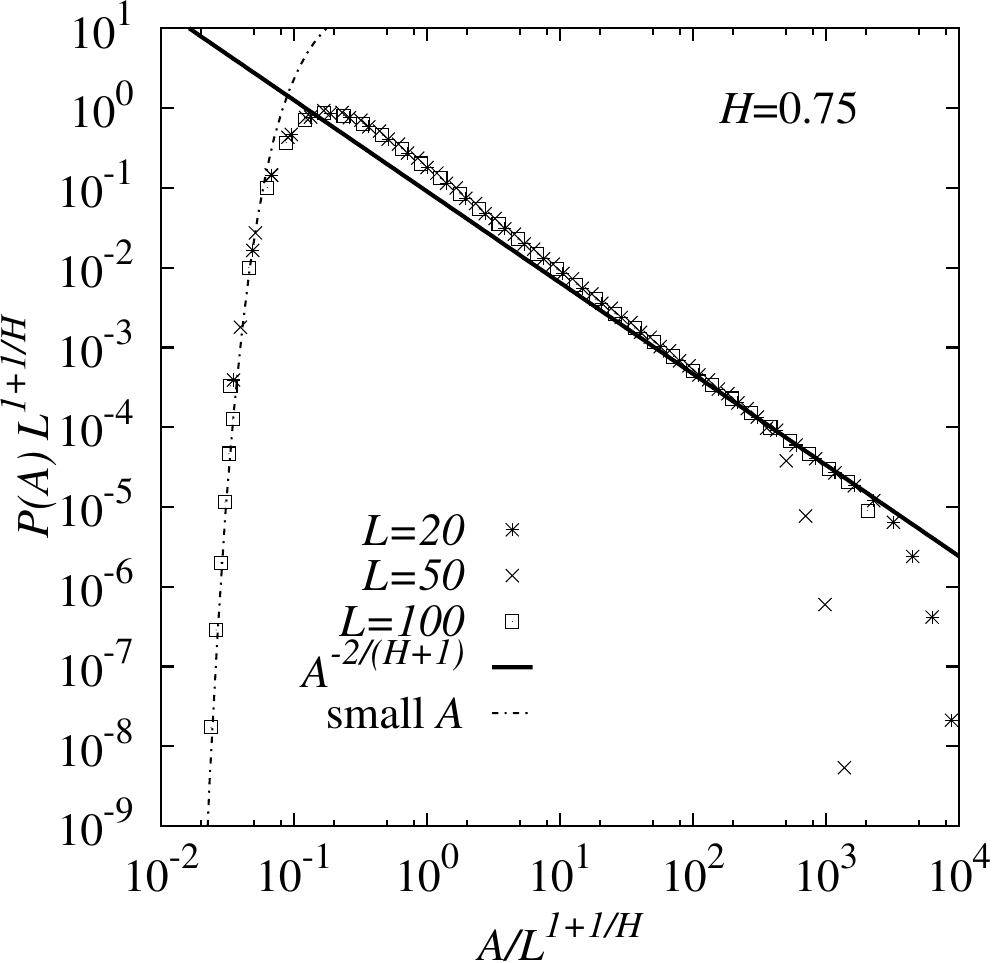}
\caption{Simulated first-passage area distributions $P(A)$ for the fractional random walk with $H=3/4$ for $L=20$, $L=50$ and $L=100$. The axes are rescaled according to our prediction~(\ref{exactscaling}) for the fBm. Also shown are two predicted asymptotics for the fBm: Eqs.~(\ref{smallA}) and (\ref{largeA}).
\label{scaledist075}}
\end{figure}

Figure \ref{scaledist075} verifies the exact scaling behavior of $P(A)$ for the fBm,  predicted by Eq.~(\ref{exactscaling}), for $H=3/4$ and three different values of $L$: $20$, $50$ and $100$. A good collapse of the three properly rescaled curves is observed. The collapsed curve describes the (presently unknown analytically) scaling function $\Phi_H(z)$, entering Eq.~(\ref{exactscaling}), for $H=3/4$. One can also see in Fig.~\ref{scaledist075} a good agreement between the distribution and the analytical predictions (\ref{smallA}) and~(\ref{largeA}), for both the left and right tails  respectively.  In the right tail the agreement is observed when $A$ is large enough, but not too large. This happens because of the finite lengths of the walks which strongly suppress very large values of $A$.

\begin{figure} [ht]
\includegraphics[width=0.40\textwidth,clip=]{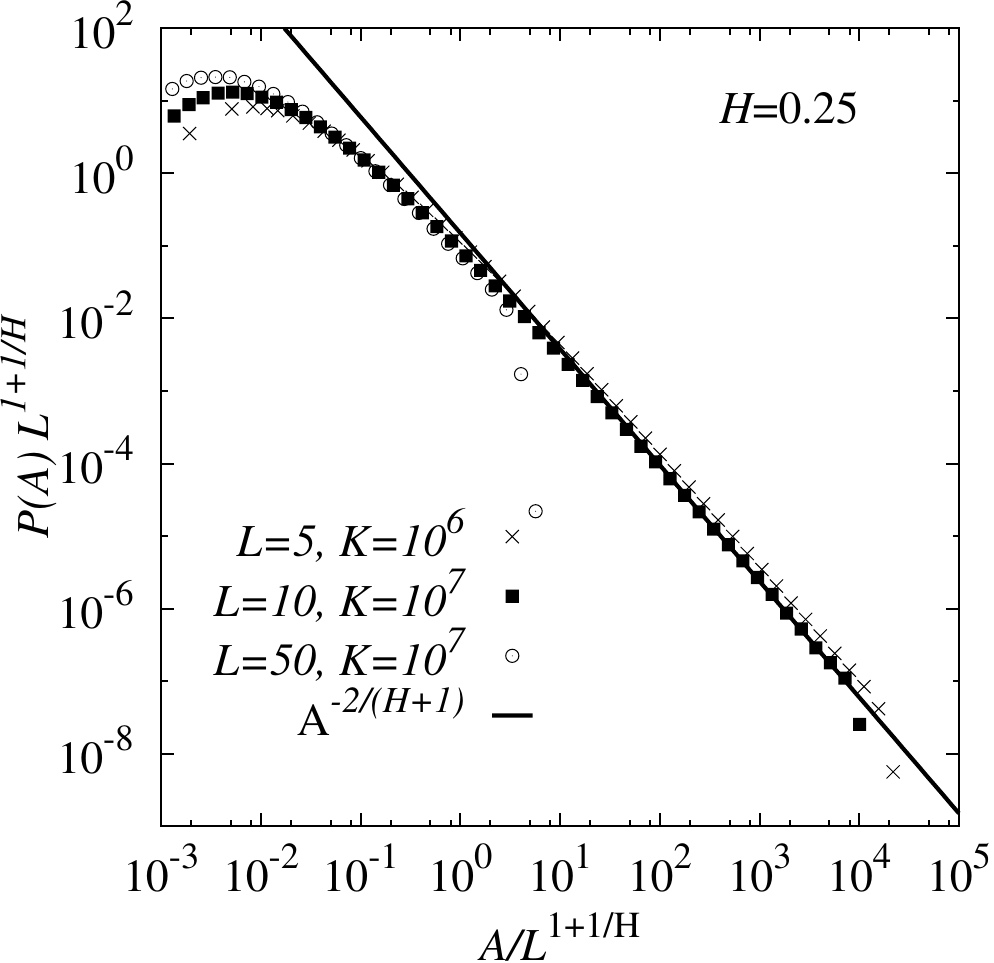}
\caption{Simulated first-passage area distributions $P(A)$ for the fractional random walk with $H=1/4$ for $L=5$, $L=10$ and $L=50$. The axes are rescaled according to our prediction~(\ref{exactscaling}) for the fBm. Also shown is the
large-$A$ tail prediction~(\ref{largeA}).
\label{scaledist025}}
\end{figure}

The corresponding results for $H=1/4$ are shown in Fig.~\ref{scaledist025}.
Since $H$ is rather small here, we have to use smaller
starting positions $L$ to obtain good statistics in the tail.
On the other hand, for the smaller $L$ that we used, the typical first-passage times
are rather small, less than 100. As a result, the approximation
of the fBm by the discrete walk becomes less accurate, in particular
near the distribution peak. This finite-size effect explains why the collapse is not as good here as in the case of $H=3/4$. It improves, however, as it should, in the large-$A$ tail. In particular, the predicted asymptotic power law decay of this tail, see Eq.~(\ref{largeA}),
is clearly visible for $L=5$ and $L=10$. For $L=50$
too many of the walks do not reach the target in the allotted time, which results in a
poor statistics and an early fall off of the tail. To significantly delay this fall off would require a prohibitively large number $K$ of steps in each run, beyond reasonable numerical effort.


\begin{figure} [ht]
\includegraphics[width=0.4\textwidth,clip=]{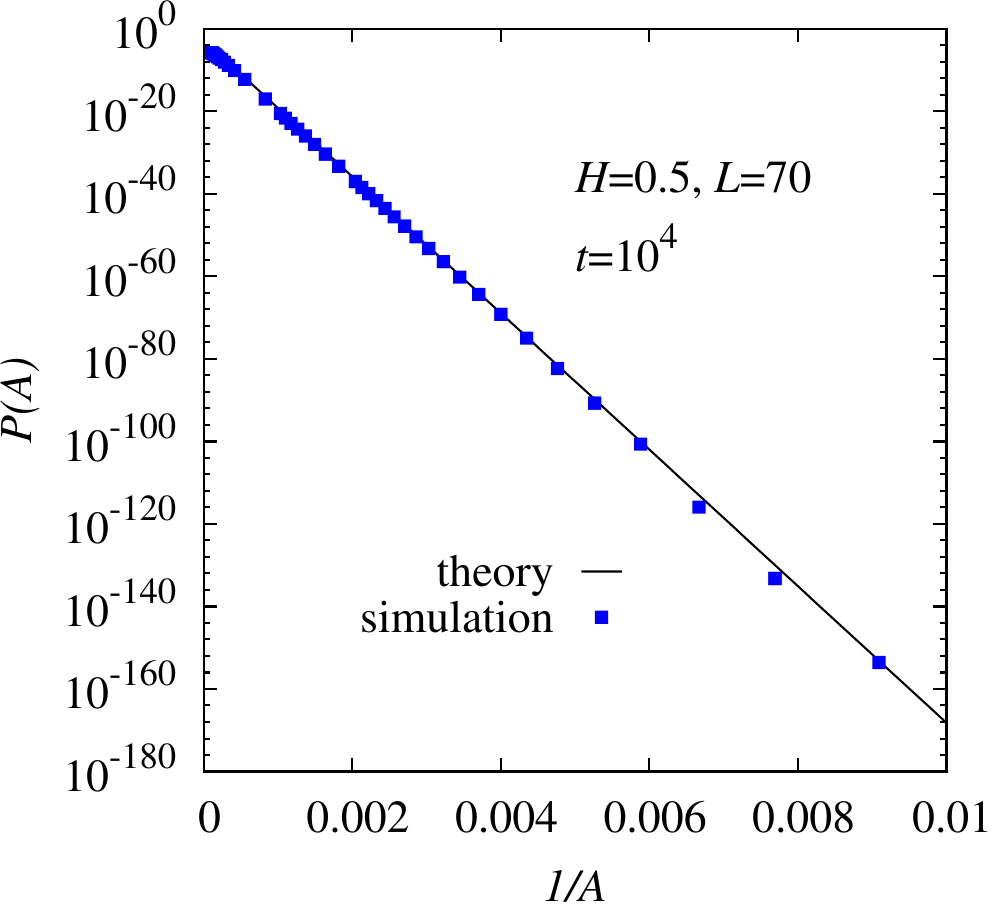}
\caption{The small-$A$ tail of the simulated first-passage area
distribution $P(A)$  as a function of $1/A$ for the fractional random walk
with $H=1/4$ (symbols).
Solid lines: theoretical prediction (\ref{smallA}) for the fBm. }
\label{smallAtail05}
\end{figure}

Now let us investigate the small-$A$ tail of $P(A)$ in more detail.
In Fig.~\ref{smallAtail05} we compare the simulated distribution
$P(A)$  for  the standard random walk ($H=1/2$) for $L=70$
with theoretical predictions~(\ref{smallA}).  As one can see, a very good agreement
is observed over more than 100 decades in probability density.
In particular, the $-\ln P\sim 1/A$ behavior, predicted by Eq.~(\ref{PAexact}), is clearly observed (notice the $1/A$ scaling of the $x$-axis). Only in the far tail a small deviation is visible. This happens at very small $A$, where the discrete character of the walk becomes very pronounced, as only a few steps of the walk occur before the first passage.

\begin{figure} [ht]
\includegraphics[width=0.4\textwidth,clip=]{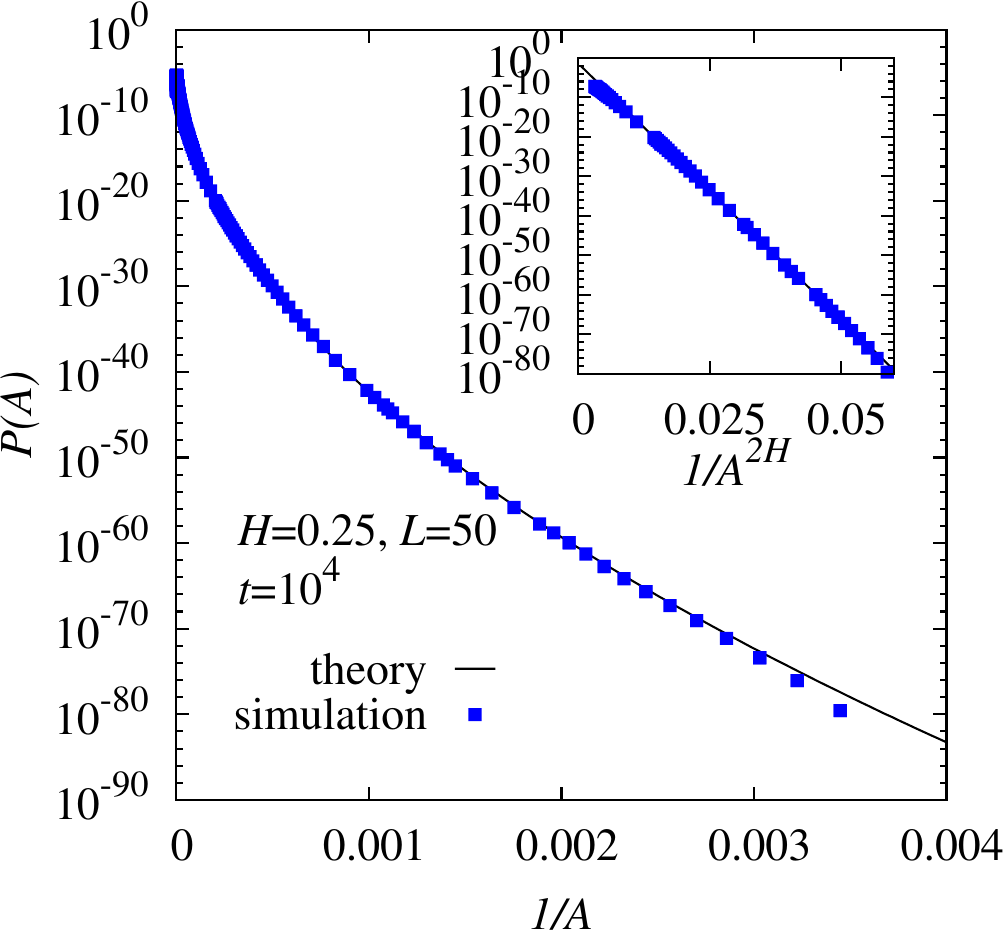}
\includegraphics[width=0.4\textwidth,clip=]{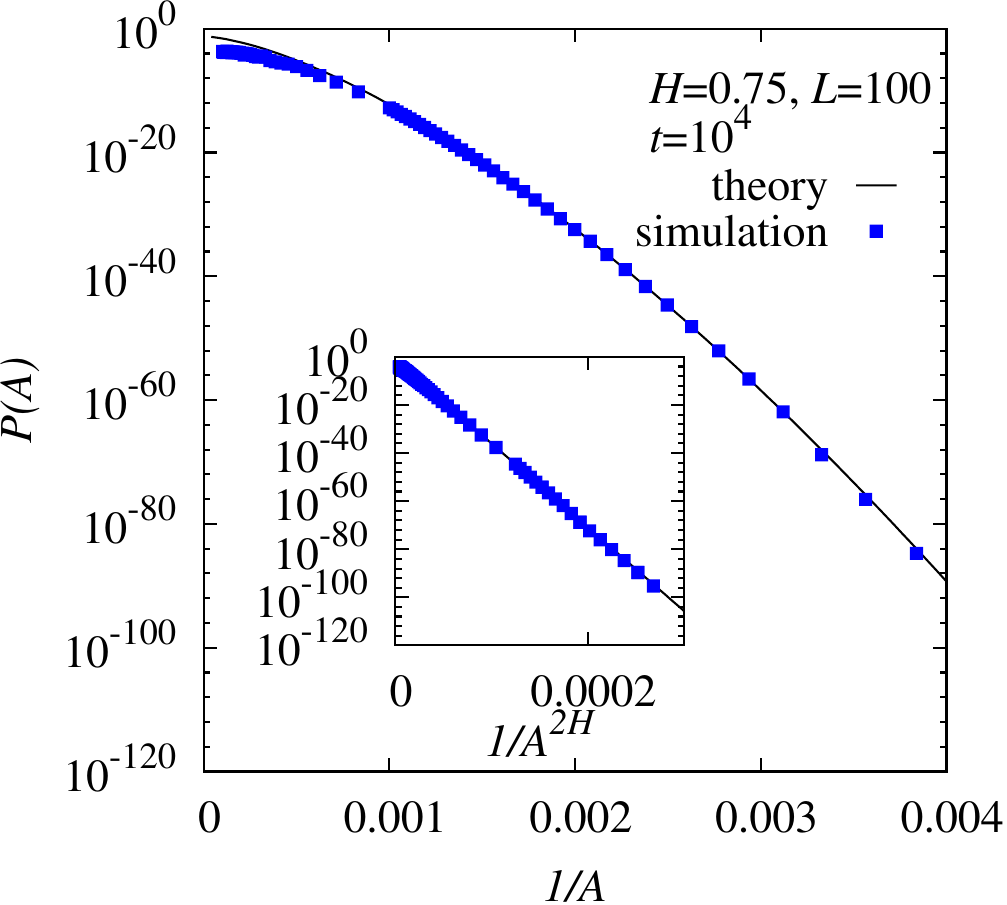}
\caption{The small-$A$ tail of the simulated first-passage area
distribution $P(A)$ as a function of $1/A$
for the fractional random walk with $H=1/4$ (top),
and $3/4$ (bottom) for indicated values of $L$ (symbols).
Solid lines: theoretical prediction (\ref{smallA}) for the fBm. The insets show the
data plotted as a function of $1/A^{2H}$ which should exhibit a straight line.
}
\label{smallAtail}
\end{figure}

In Fig.~\ref{smallAtail} we show the corresponding results for
$H=1/4$ with $L=50$  and $H=3/4$ with $L=100$.
Again a very good agreement is observed, confirming the analytical predictions.
The expected $-\ln P \sim A^{-2H}$, see Eq.~(\ref{smallA}), is clearly visible in the insets. 

\subsection{Optimal paths}

\begin{figure} [ht]
\includegraphics[width=0.4\textwidth,clip=]{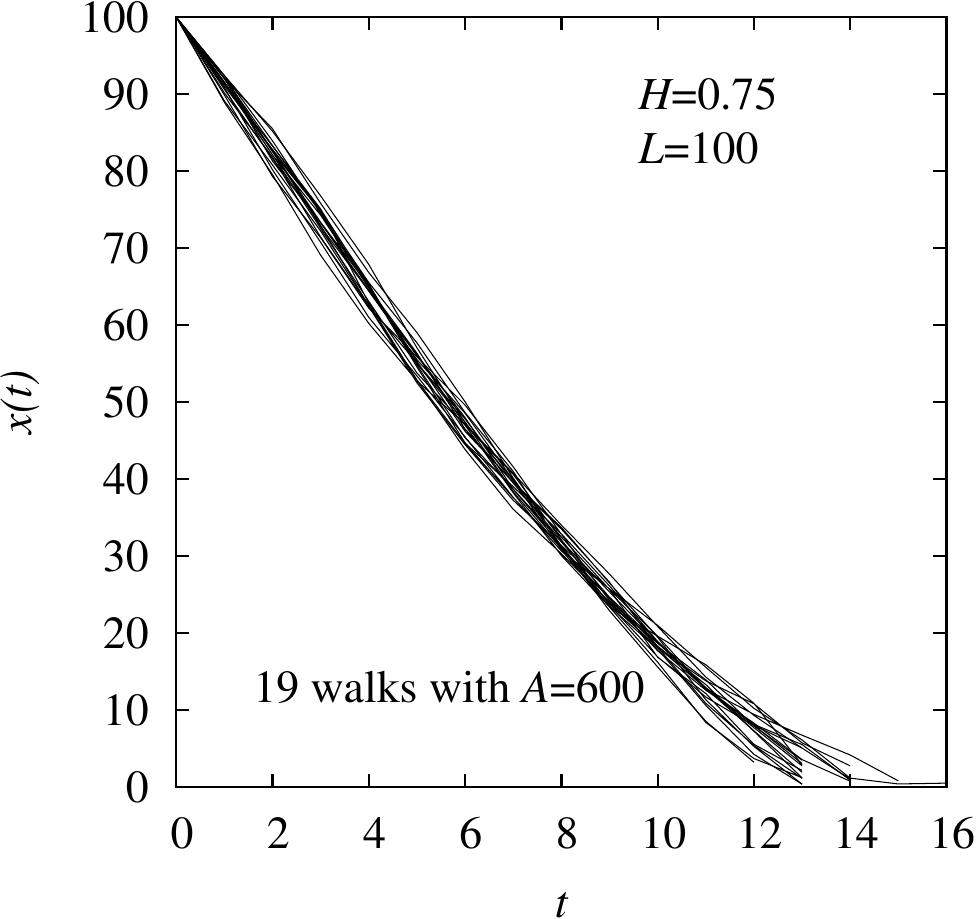}
\includegraphics[width=0.4\textwidth,clip=]{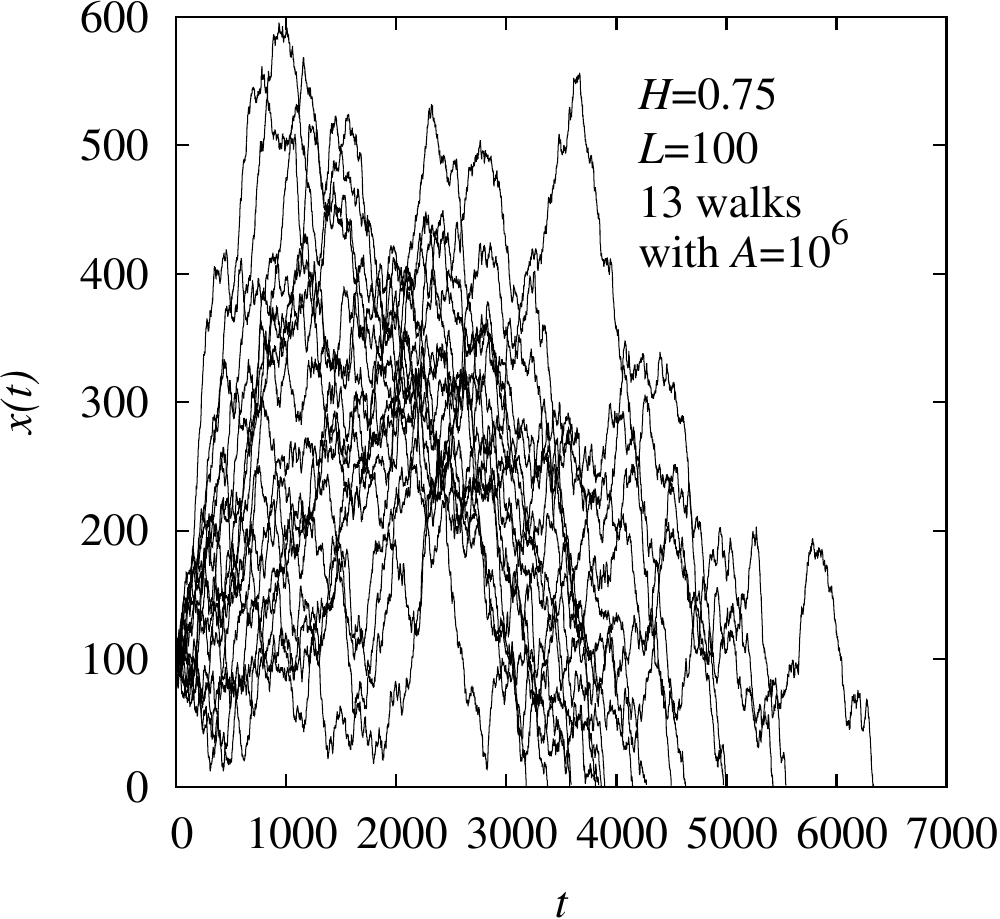}
\caption{Several sample paths $x(t)$ for $H=3/4$, $L=100$ corresponding to very small
area $A=600$ (top) and very large area $A=10^6$ (bottom), respectively.
The typical area for these $H$ and $L$ is $A\sim 10^4$.}
\label{fig:sample:paths}
\end{figure}

Next, we analyze the paths conditioned on specific values of $A$. To remind the reader, theory predicts that the probability of observing unusually small values of $A$ is dominated by a well-defined single optimal path. On the contrary, unusually large values of $A$ can come from multiple different paths.

To analyze the paths, we occasionally stored during the simulations
the full paths along with the corresponding value of the first-passage
area. This allowed us to select paths conditioned on specific values, \textit{i.e.}
small intervals, of $A$.

In Fig.~\ref{fig:sample:paths} we consider the case $H=3/4$ with $L=100$.
For these parameters, typical values of $A$ are about $10^4$: see
Fig.~\ref{scaledist075}, where the distribution peak is located near
$A/L^{1+1/H}\approx 0.2$. Therefore, as representatives of the small-$A$
tail behavior, we selected the paths exhibiting $A=600$,
actually $A\in [600,601)$. This resulted in 19 paths which are all
shown in the top figure. As one can see, these paths are very close to each other.
In particular, the first passage times $T$ do not differ much.
On the contrary, for large values of $A\approx 10^6$, \textit{i.e.}
$A\in[9.95\time 10^5,10.05\times 10^5]$, the sampling resulted in
 13 paths which are shown in the bottom figure. These paths are strikingly
different. In particular, they exhibit a broad distribution of
the first-passage times $T$.

\begin{figure} [ht]
\includegraphics[width=0.40\textwidth,clip=]{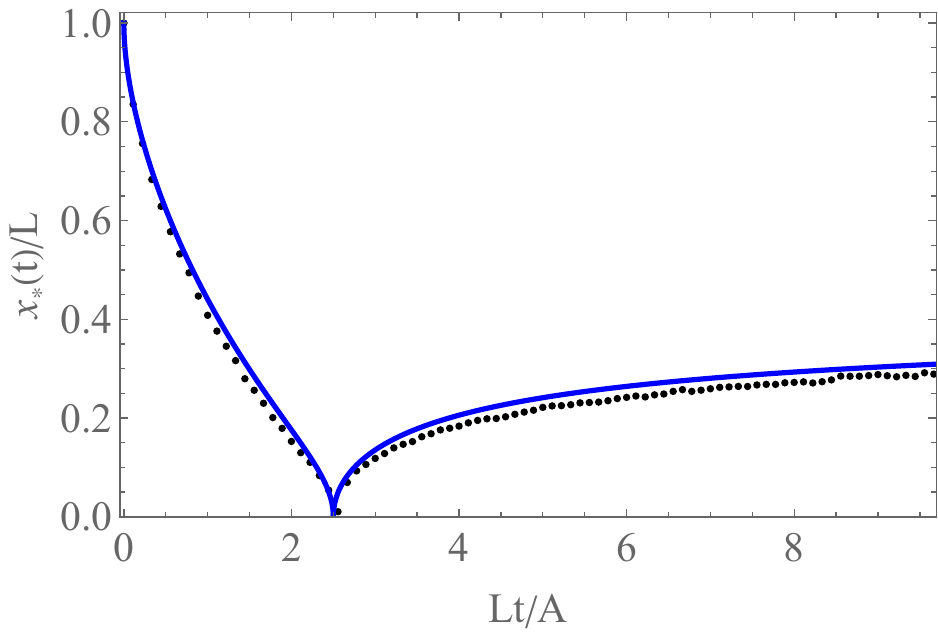}
\includegraphics[width=0.40\textwidth,clip=]{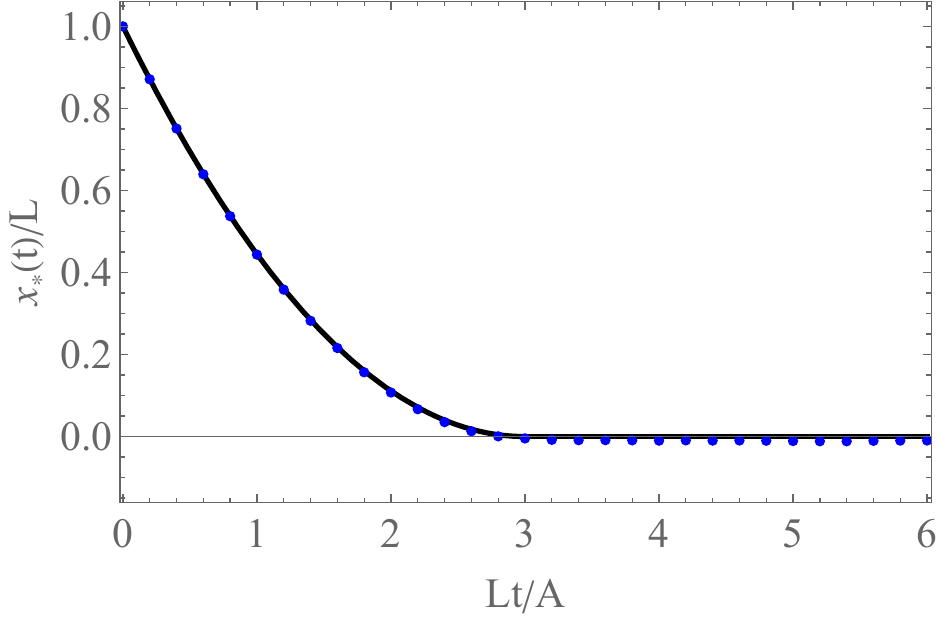}
\includegraphics[width=0.40\textwidth,clip=]{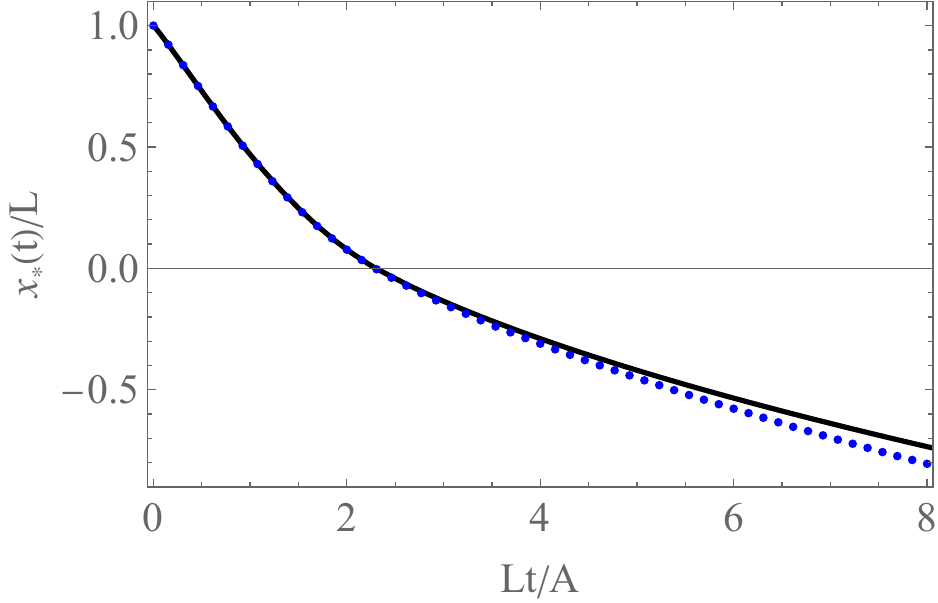}
\caption{The symbols show the average of simulated
paths $x(t)$ for $H=1/4$ and $L=150$
conditioned on $A=450$ (top), for $H=1/2$ and $L=70$ conditioned on
$A=350$ (middle), and for $H=3/4$ and $L=100$ conditioned on $A=650$ (bottom).
The line represents the analytically obtained optimal paths $x_*(t)$ from Ref. \cite{MO2022}.
The $x$-coordinate is rescaled by $L$, time is rescaled by  $A/L$
(the latter is proportional to the optimal first-passage
time \cite{MO2022}).}
\label{opath}
\end{figure}

For the small values of $A$, where the paths are close to each other,
we averaged them.  Figure \ref{opath} presents, for $H=1/4$,
$1/2$ and $3/4$, the average  simulated  paths $x(t)$
constrained on sufficiently small $A$.  Also shown are theoretical
predictions for the optimal paths $x(t)$ with these values of $H$ from Ref. \cite{MO2022}. The
$x$-coordinate is rescaled by $L$, and time is rescaled by $A/L$.  As
one can see, the simulations and theory agree quite well. Importantly,
the agreement persists beyond the time interval $0<t<T$: to the
``future" where, at $H\neq 1/2$, the optimal path is
still non-trivial due to the non-Markov nature of the fBm \cite{MO2022}.

For $H<1/2$ we observe the remarkable phenomenon of
reflection of the path from the origin, predicted in
Ref. \cite{MO2022}. The reflection exhibits a cusp singularity at the
reflection point.  Finally, in all three cases, the measured
first-passage time $T$ is close to the theoretically predicted
\emph{optimal} first-passage time $T_*=2A(1+H)/L$ for $H\leq 1/2$, and
$T_*=A(1+H)/HL$ for $H\geq 1/2$ \cite{MO2022}.

\section{Statistics of $A_n$}

One can also study the statistics of a more general first-passage fractional Brownian functional, which has the form
\begin{equation}\label{fpfunctional}
A_n=\int_0^T x^n(t) \,dt\,,
\end{equation}
where $T$ is again the first-passage time. As of present, $P(A_n)$ --  the probability distribution of $A_n$ at given $x(0)=L$ -- is known exactly (for $n>-2$) only for the standard Brownian motion $H=1/2$ \cite{MM2020a}. The particular case $n=1$ corresponds to the area under $x(t)$ that we have been dealing with in this paper, whereas $n=0$ corresponds to the statistics of the first-passage time itself. Additional values of $n$ can be of interest as well, as is the case of the standard Brownian motion \cite{MM2020a}.

Some important properties of  $P(A_n)$ can be established immediately. Indeed, dimensional analysis 
yields the following exact
scaling behavior of $P(A_n)$:
\begin{equation}\label{exactscalingn}
P(A_n) = \frac{D^\frac{1}{2H}}{L^{n+\frac{1}{H}}}\,\Phi^{(n)}_H\left(\frac{D^\frac{1}{2H} A_n}{L^{n+\frac{1}{H}}}\right)\,,
\end{equation}
which generalizes Eq. ~(\ref{exactscaling}). The presently unknown scaling function $\Phi^{(n)}_H(z)$ depends, apart from $z$, also on $H$ and $n$.

The structure of the $A_n\to 0$ tail of $P(A_n)$ can be predicted from a geometrical-optics argument. Indeed, at very small $A_n$, the probability distribution $P(A_n)$ is expected to be exponentially small and exhibit the characteristic $1/D$ weak-noise scaling inside the exponent. Then, using the scaling relation (\ref{exactscalingn}), we arrive at the tail
\begin{equation}\label{smallAn}
P(A_n\to 0) \sim \exp\left[-\sigma_n(H) \frac{L^{2+2nH}}{DA^{2H}}\right]\,.
\end{equation}
which generalizes Eq.~(\ref{smallAn}).  The character of essential singularity at $A\to 0$, predicted by Eq.~(\ref{smallAn}), is determined by $H$. Interestingly, it is independent of $n$. For $H=1/2$ the latter feature was already observed in Ref. \cite{MM2020a}. In order to determine the presently unknown dimensionless factor $\sigma_n(H)$, one should solve the geometrical-optics problem, that is minimize the action for the fBm subject to constraint (\ref{fpfunctional}), the boundary conditions $x(0)=L$ and $x(T)=0$, and the inequality $x(0<t<T)>0$. The minimization should be done with respect to the path $x(t)$ and the first-passage time $T$.

In its turn, the same scaling arguments, which led us to the prediction of the large-$A$ tail (\ref{largeA}) of $P(A)$, can be used to predict the large-$A_n$ tail of $P(A_n)$:
\begin{equation}\label{largeAn}
P(A_n\to \infty) \sim \frac{L^{\frac{1}{H}-1}}{D^{\frac{(1-H) (nH-H+1)}{2 H (nH +1)}}A_n^{1+\frac{1-H}{nH+1}}}\,.
\end{equation}
For all $n>-1/H$ and all $H\in (0,1)$ this tail is fat: the exponent of $A_n$  is greater than $1$. Therefore, the average value of $A_n$ diverges.

As to be expected, Eqs.~(\ref{exactscalingn})-(\ref{largeAn}) coincide with Eqs.~(\ref{exactscaling}), (\ref{smallA}) and (\ref{largeA}), respectively,  for $n=1$. For $n=0$
Eq.~(\ref{largeAn}) recovers the large-$T$ tail~(\ref{flargeA}) of the first-passage time distribution.

\section{Summary and Discussion}

In the present work, we have studied the distribution $P(A)$ of
the first-passage area for the fBm.
We started with establishing exact scaling behavior of $P(A)$, see Eq.~(\ref{exactscaling}). The small-$A$ asymptotic
of $P(A)$ was predicted earlier \cite{MO2022}. Here we
determined the fat-tail large-$A$ asymptotic behavior of $P(A)$,
see Eq.~(\ref{largeA}).

The main effort of this work  was to study
$P(A)$ numerically, for values of $H$ in the subdiffusive ($H<1/2$)
and superdiffusive ($H>1/2$) regions,
as well as for the standard Brownian case ($H=1/2$).
For this purpose, we approximated the
fBm by correlated discrete-time random walks.
For typical and atypically large area $A$, we have employed
simple sampling and observed
a good agreement between the theory and simulations in the right tail of $P(A)$.
In addition, by applying large-deviation
approaches, we have studied the small-$A$ tail with high precision down to
probability densities as small as $10^{-190}$. Here we have observed a very
good agreement between analytical predictions of Ref. \cite{MO2022}. We also
verified in simulations the exact scaling behavior of $P(A)$, described by Eq.~(\ref{exactscaling}).

Furthermore, by storing sampled configurations
of the simulated walks along with the values of $A$, we were able
to study the shape of the walks
conditioned on the area $A$. For larger than typical values of $A$,
the shape of the walks and the corresponding first-passage times
fluctuate a lot from sample to sample, even if the walks exhibit the
same area. This is what to be expected in a regime where the proper
path integral of the process \cite{MBO2022} is contributed to by multiple, although unusual,
paths.  On the contrary, for unusually small values of $A$, the walks look very
much alike, and the averaged walks agree very well with the optimal paths
previously obtained analytically \cite{MO2022}. Again, this is to be expected in a regime
where the path integral is dominated by the well defined optimal path.

We have also shown that our results can be extended to the statistics of a more general first-passage fractional Brownian functional (\ref{fpfunctional}). It would be interesting to continue this line of study and, in particular, calculate the function $\sigma_n(H)$ which enters Eq.~(\ref{smallAn}).

In the absence of exact result for the scaling function $\Phi_H(z)$, it would be 
desirable to access next-order corrections to the leading-order small-$z$ and large-$z$ 
asymptotics that we considered here. In particular, this would improve agreement between
the finite-size simulations and analytical predictions. The next-order correction to
the small-$A$ asymptotic (\ref{smallA}) would require going beyond
the saddle-point approximation to the path integral for the fBm, 
which was used in Ref.~\cite{MO2022}. 

For the large-$A$ asymptotic (\ref{largeA}), the situation is not better: here we do
not know the amplitude of the power law -- a factor  $O(1)$ which depends only on $H$. It is unclear to us how to calculate it analytically. We hope that this work 
will stimulate further studies.

\section*{Acknowledgments}
We are grateful to S. N. Majumdar and G. Oshanin for useful discussions.
The authors were financially
supported by the Erwin Schrödinger International Institute
for Mathematics and Physics (Vienna, Austria) within the Thematic Programme
``Large Deviations, Extremes and Anomalous Transport in Non-equilibrium
Systems'' during which part of the work was performed.
The simulations were performed at the HPC Cluster
CARL, located at the University of Oldenburg (Germany)
and funded by the DFG through its Major
Research Instrumentation Program (INST 184/157-
1 FUGG) and the Ministry of Science and Culture
(MWK) of the Lower Saxony State. B. M. was supported by the Israel
Science Foundation (Grant No. 1499/20).



\begin{thebibliography}{99}
\bibitem{Kolmogorov} A. N. Kolmogorov, CR (Dokl.) Acad. Sci.
URSS \textbf{26}, 115 (1940).
\bibitem{Mandelbrot} B. B. Mandelbrot and J. W. van Ness, SIAM Review \textbf{10}, 422 (1968).
\bibitem{Krug} J. Krug, H. Kallabis, S. N. Majumdar, S. J. Cornell, A. J. Bray,and C. Sire,
Phys. Rev. E \textbf{56}, 2702 (1997).

\bibitem{weiss} M. Weiss,
Phys. Rev. E {\bf 88}, 010101(R) (2013).
\bibitem{weiss2} D. Ernst, M. Hellmann, J. K\"ohler, and M. Weiss,  Soft Matter {\bf 8}, 4886
(2012).
\bibitem{weber} S. C. Weber, A. J. Spakowitz, and J. A. Theriot, Phys. Rev. Lett. {\bf 104}, 238102 (2010).

\bibitem{garini} I. Bronshtein, E. Kepten, I. Kanter, S. Berezin, M. Lindner,
A. B. Redwood, S. Mai, S. Gonzalo, R. Foisner, Y. Shav-Tal, and Y. Garini,
Nat. Commun. {\bf 6}, 8044 (2015).

\bibitem{prx} D. Krapf, N. Lukat, E. Marinari, R. Metzler, G. Oshanin, C. Selhuber-Unkel,
A. Squarcini, L. Stadler, M. Weiss,  and X. Xu,
Phys. Rev. X {\bf 9}, 011019 (2019).

\bibitem{vojta} S. Janusonis, N. Detering, R. Metzler, and T. Vojta,
Frontiers Comp. Neurosci. {\bf 14}, 56 (2020).





\bibitem{Walter} J.-C. Walter, A. Ferrantini, E. Carlon, and C. Vanderzande,
Phys. Rev. E \textbf{85}, 031120 (2012).

\bibitem{Amitai} A. Amitai, Y. Kantor, and M. Kardar, 
Phys. Rev. E \textbf{81}, 011107 (2010).

\bibitem{Zoia} A. Zoia, A. Rosso, and S. N. Majumdar, 
Phys. Rev. Lett. \textbf{102}, 120602 (2009).

\bibitem{Dubbeldam} J. L. A. Dubbeldam, V. G. Rostiashvili, A. Milchev, and
T. A. Vilgis, 
Phys. Rev. E \textbf{83}, 011802 (2011).

\bibitem{Palyulin} V. Palyulin, T. Ala-Nissila, and R. Metzler, 
Soft Matter \textbf{10}, 9016 (2014).

\bibitem{Kukla} V. Kukla, J. Kornatowski, D. Demuth, I. Girnus, H. Pfeifer,
L. V. C. Rees, S. Schunk, K. K. Unger, and J. K¨arger,
Science \textbf{272}, 702 (1996).

\bibitem{Wei} Q.-H. Wei, C. Bechinger, and P. Leiderer, 
Science \textbf{287}, 625 (2000).

\bibitem{chanel} O. B\'enichou, P. Illian, G. Oshanin, A. Sarracino, and R. Voituriez,
J. Phys.: Condens. Matter  {\bf 30}, 443001 (2018).

\bibitem{ralf} R. Metzler, J.-H. Jeon, A. G. Cherstvy, and E. Barkai,
 Phys. Chem. Chem. Phys. {\bf 16}, 24128 (2014).

\bibitem{KM2005} M. J. Kearney and S. N. Majumdar,  
J. Phys. A: Math. Gen. \textbf{38}, 4097 (2005).

\bibitem{MM2020a}  S. N. Majumdar and B. Meerson, J. Stat.
Mech. (2020) 023202. Corrigendum: J. Stat. Mech. (2021) 039801.



\bibitem{DR1989} D. Dhar and R. Ramaswamy, 
Phys. Rev. Lett. {\bf 63}, 1659 (1989).

\bibitem{K2004} M. J. Kearney, 
J. Phys. A: Math. Gen. {\bf 37}, 8421 (2004).

\bibitem{PB1995} T. Prellberg and R. Brak, 
J. Stat. Phys. {\bf 78}, 701 (1995).

\bibitem{C2002} C. Richard, 
J. Stat. Phys.
{\bf 108}, 459 (2002).

\bibitem{MK2007} S. N. Majumdar and M. J. Kearney, 
Phys. Rev. E {\bf 76}, 031130 (2007).


\bibitem{Barenblatt} G. I. Barenblatt, \textit{Scaling, Similarity, and Intermediate Asymptotics} (Cambridge University Press, Cambridge, UK, 1996).

\bibitem{MO2022} B. Meerson and G. Oshanin, Phys. Rev. E \textbf{105}, 064137 (2022).
\bibitem{MBO2022}  B. Meerson, O. B\'{e}nichou, and G. Oshanin, Phys. Rev. E \textbf{106}, L062102 (2022).

\bibitem{Hansen} A. Hansen, T. Engoy, and K. J. Maloy, Fractals \textbf{2} 527, (1994).
\bibitem{Maslov} S. Maslov, M. Paczuski, and P. Bak, Phys. Rev. Lett. \textbf{73}, 2162 (1994).

\bibitem{Ding} M. Ding and W. Yang, Phys. Rev. E \textbf{52}, 207 (1995).

\bibitem{Krugetal} J. Krug, H. Kallabis, S. N. Majumdar, S. J. Cornell, A. J. Bray, and C. Sire, Phys. Rev. E \textbf{56}, 2702 (1997).


\bibitem{Molchan} G. M. Molchan, Theory Probab. Appl. \textbf{44}, 97 (1999).

\bibitem{Aurzada} F. Aurzada, Elect. Comm. in Probab. \textbf{16}, 392 (2011).


\bibitem{davies1987} R. B. Davies and D. S. Harte, Biometrika \textbf{74},
95 (1987).

\bibitem{wood1994}  A. T. A. Wood and G.  Chan, J. Comp. Graph. Stat.
\textbf{3}, 409 (1994).

\bibitem{dietrich1997}
 C. R. Dietrich and G. N. Newsam, SIAM J. Sci. Comput., \textbf{18},
1088 (1997).


\bibitem{fBm_MC2013}
  A.K. Hartmann,  S. N. Majumdar, and A. Rosso,
  Phys. Rev. E \textbf{88}, 022119 (2013).

\bibitem{gsl2006}
  M. Galassi,  J. Davies, J. Theiler, B. Gough,
 G. Jungman, M. Booth, and F. Rossi,
  \textit{GNU Scientific Library Reference Manual},
  (Network Theory Ltd., 2006).


\bibitem{walter2020}
B. Walter and K. J. Wiese, Phys. Rev. E \textbf{101}, 043312 (2020).


\bibitem{hammersley1956} J. M. Hammersley and K. W. Morton,
Math. Proc. Cambr. Phil. Soc \textbf{52}, 449 (1956).

\bibitem{dellago1998}
D. Dellago, P. G.  Bolhuis, F. S. Csajka  and D. Chandler,
J. Chem. Phys. \textbf{108}, 1964 (1998).

\bibitem{crooks2001}
G. E. Crooks and D. Chandler, Phys. Rev. E \textbf{64}, 026109 (2001).

\bibitem{hartmann2004}
A. Engel, R. Monasson, and A. K.  Hartmann, J. Stat. Phys. textbf{117}
387 (2004).

\bibitem{hartmann2011}
A. K. Hartmann,
Eur. Phys. J B. \textbf{84}, 627 (2011).

\bibitem{hartmann2017}
A. K. Hartmann,
Eur. Phys. J. Spec. Top. \textbf{226}, 567 (2017).

\bibitem{schawe2019}
H. Schawe and A. K. Hartmann,
Eur. Phys. J. B. \textbf{92}, 73 (2019).

\bibitem{Dewenter_2015}
T. Dewenter and A. K.  Hartmann,
New Journal of Physics \textbf{17}, 015005 (2015).

\bibitem{feld19}
Y. Feld Y and A. K. Hartmann Chaos \textbf{29}, 113103 (2019).

\bibitem{claussen2015}
G. Claussen, A. K.  Hartmann and S. N. Majumdar,
Phys. Rev. E. \textbf{91}, 052104 (2015).

\bibitem{Schawe2019No2}
H. Schawe and A.K. Hartmann,
J. Physics: Conf. Ser. \textbf{1290}, 012029 (2019).

\bibitem{schawe2020}
H. Schawe and A.K.  Hartmann,
Phys. Rev. E. \textbf{102}, 062141 (2020).

\bibitem{Chevallier2020}
A. Chevallier and F. Cazals,
J. Comput. Phys. \textbf{410}, 109366 (2020).

\bibitem{Koerner2006}
M. Körner and H. G.  Katzgraber, and A. K. Hartmann,
JSTAT \textbf{2006}, P04005 (2006).

\bibitem{boerjes2019}
J. B\"orjes, H. Schawe, and A. K.  Hartmann,
Phys. Rev. E. \textbf{99}, 042104 (2019).

\bibitem{krabbe2020}
P. Krabbe, H. Schawe, and A. K. Hartmann,
Phys. Rev. E. \textbf{101}, 062109 (2020).





\bibitem{Hartmann2018} A. K. Hartmann, P. Le Doussal, S. N. Majumdar, A. Rosso,
and G. Schehr, Europhys. Lett. \textbf{121}, 67004 (2018).

\bibitem{HMS1}
A. K.  Hartmann, B. Meerson, and P. Sasorov,
Phys. Rev.  Res. \textbf{1}, 032043(R) (2019).

\bibitem{HMS2} A.K. Hartmann, B. Meerson, and P. Sasorov,
Phys. Rev. E \textbf{104}, 054125 (2021).

\bibitem{bucklew2004}
  J. A. Bucklew, \textit{Introduction to rare event simulation},
  (Springer-Verlag, New York, 2004).


\bibitem{newman1999} M.E.J. Newman and G. T. Barkema,
Monte Carlo Methods in Statistical Physics, Clarendon Press (Oxford), 1999.




\bibitem{work_ising2014} A.K. Hartmann, Phys. Rev. E {\bf 89}, 052103 (2014).


\bibitem{align2002} A.K. Hartmann, Phys. Rev. E {\bf 65}, 056102 (2002).

\end{thebibliography}
\end{document}